\documentclass[a4paper]{PoS}

\title{Einstein Probe --- a small mission to monitor and explore the dynamic X-ray Universe}

\ShortTitle{Einstein Probe}

\author{
 \speaker{Weimin Yuan}$^a$\thanks{E-mail:wmy@nao.cas.cn},  
C. Zhang$^a$, H. Feng$^b$, S.-N. Zhang$^{c}$, Z.-X. Ling$^a$, D. Zhao$^a$, J. Deng$^a$,
Y. Qiu$^a$, 
        J.P. Osborne$^d$,
        P. O'Brien$^d$, 
        R. Willingale$^d$, J. Lapington$^d$, 
        G.W. Fraser$^d$\footnote{Deceased, March 2014},
        \&  the Einstein Probe team\\
        \llap{$öa$} Key Laboratory of Space Astronomy and Technology, National Astronomical Observatories (NAOC), Chinese Academy of Sciences, Datun Rd 20A, Beijing 100012 China\\
         \llap{$öb$} Department of Engineering Physics, Tsinghua University, Beijing 100084, China\\
         \llap{$öc$} Institute of High Energy Physics (IHEP), Chinese Academy of Sciences, 100049, Beijing, China\\
         \llap{$öd$} Department of Physics and Astronomy, University of Leicester, Leicester, LE1 7RH, UK     
}

\abstract{
Einstein Probe  is a  small  mission dedicated to time-domain high-energy astrophysics.
Its primary goals are to discover high-energy  transients and to monitor variable objects 
in the 0.5--4\,keV X-rays, 
at higher sensitivity  by one order of magnitude than those of the ones currently in orbit.
Its wide-field imaging capability, 
featuring a large instantaneous field-of-view ($60^\circ \times60^\circ$, $\sim1.1$\,sr),
 is achieved by using established technology of micro-pore (MPO) lobster-eye optics, 
 thereby offering unprecedentedly high sensitivity  and large Grasp 
 (effective area times field-of-view).
To complement this powerful monitoring ability, 
it also carries a narrow-field, sensitive follow-up X-ray telescope 
based on the same MPO technology to perform follow-up observations of newly-discovered transients. 
Public transient alerts will be downlinked rapidly, 
so as to trigger multi-wavelength follow-up observations from the world-wide community.
Over three of its 97-minute orbits almost the entire night sky will be sampled, 
with cadences ranging from 5 to 25 times per day. 
The scientific objectives of the mission are: 
to discover otherwise quiescent black holes over all astrophysical mass scales by 
 detecting their rare X-ray transient flares, particularly tidal disruption of stars by 
 massive black holes at galactic centers;
 to detect and precisely locate the electromagnetic sources of gravitational-wave transients;
 to carry out systematic surveys  of X-ray transients and characterise
 the variability of  X-ray sources,  such as high-redshift gamma-ray bursts, 
 supernova shock breakouts, 
  X-ray binaries of compact objects, gamma-ray bursts, active galactic nuclei and stellar coronal flares, etc. 
Einstein Probe 
has been selected as a candidate mission of priority (no further selection needed)
in the Space Science Programme of 
the Chinese Academy of Sciences, aiming for launch around 2020.
}

\FullConference{Swift: 10 Years of Discovery,\\
		2-5 December 2014\\
		La Sapienza University, Rome, Italy }

\begin{document}

\section{Introduction}

Transients and variable objects pervade the X-ray Universe; some indeed are  
the most energetic outbursts in the Universe. 
Observations and interpretations of these phenomena have revolutionised 
our understanding of the Universe and its underlying physical laws. 
They provide otherwise inaccessible laboratories to explore the limits of contemporary physics 
and to study matter under extreme conditions. 
A rich variety of transients and variables, with diverse timescales from sub-seconds to years, have been discovered and extensively studied since the early days of X-ray astronomy, 
thanks to successive all-sky monitors in the X-rays and $\gamma$-rays.
This exciting field of discovery is  highlighted by \emph{Swift} \cite{geh04},
which has made many remarkable and surprising discoveries during the past 10-year's operation,
from short gamma-ray bursts (GBRs) and high-redshift GRBs, to
supernova shock breakouts, magnetars, and relativistic tidal disruption events, etc.\ \cite{geh15}.
More discoveries continue to be made in recent years with the 
joining of \emph{Fermi} \cite{fermi} and \emph{MAXI} \cite{mats09} in the game. 

Despite  rapid progress in this field, many key questions remain unanswered, 
and yet more have been introduced.
New phenomena continue to be discovered and appeal for observations on a large scale for characterisation, others are highly expected and await discovery. 
Particularly, with the advent of major facilities across the electromagnetic spectrum 
and in the multi-messenger realms of gravitational wave and neutrinos,
time-domain astronomy will enter a golden era towards the end of this decade. 
This makes  monitoring and exploration of the X-ray sky even more exciting, 
and calls for the next generation of wide-field 
monitoring instruments with  higher
sensitivity so as to extend  the horizons of the monitored Universe
significantly beyond the reach of the  ones currently 
in orbit.

Driven by the great scientific potential, 
the Einstein Probe (EP) \footnote{\it{http://ep.bao.ac.cn}} mission was proposed by the 
Chinese Academy of Sciences (CAS) institutes (NAOC \& IHEP) and Tsinghua University in 2012, 
in response to a call of the CAS for small- and medium-sized candidate missions 
in its newly initiated space science  programme. 
In 2015, EP has been selected to be among three candidate missions of priority 
aiming for launch around 2020, and is funded for advanced study by the CAS 
under its {\it Strategic Priority Research Programme of Space Science}.
The project is open for international participation via collaboration, 
which currently involves mainly the UK.

\section{Scientific Objectives}
Are black holes prevalent in our Universe and how does matter fall onto them?  
What kind of electromagnetic radiation and sources are 
associated with cosmic gravitational-wave events, 
and where and how are they both produced? 
When and where did the first generation of stars form and how did they re-ionize the dark early Universe? What happens at the very first instance when a star explodes and generates a supernova? 
EP will try to address these questions by capturing faint flashes of X-ray radiation produced by energetic events within a cosmic horizon far beyond the reach of any current and previous missions. Its primary scientific objectives are:

\noindent 
(1) Reveal quiescent black holes at almost all astrophysical mass scales and study how matter falls onto them by detecting  transient X-ray flares, particularly stars being tidally-disrupted by otherwise dormant massive black holes at galactic centres.\\
(2) Discover the X-ray photonic counterparts of gravitational-wave transients 
 found with the next generation of gravitational-wave detectors and  precisely locate them.\\
(3) Carry out systematic and sensitive surveys of high-energy transients, to
discover faint X-ray transients, such as  high-redshift GRBs, supernova
shock breakout, and new types of  transients.

\subsection{Discover and explore otherwise quiescent black holes via catching X-ray flares}

A star approaching a massive black hole (MBH) closely enough 
will be tidally disrupted and accreted 
when its self-gravity cannot balance the strong tidal force,
producing a flare of electromagnetic radiation \cite{rees88}. 
First discovered in the ROSAT survey \cite{kb99} and 
later from archival multiwavelength surveys,
the majority of such tidal disruption events (TDEs) identified so far 
(about two dozen) 
were found via their luminous transient X-ray emission \cite{kom12}.
As perhaps the most unique signature 
of the existence of MBHs in otherwise quiescent galactic nuclei \cite{rees88}, 
TDEs provide a census of MBHs in the Universe, 
i.e.\ to constrain their occupation fractions in various types and masses of galaxies,
which is essential for understanding the formation and evolution 
of galaxies and MBHs. 
Furthermore, observations of the detailed processes of these violent  events 
would provide otherwise inaccessible laboratories to explore how, 
and what happens when, matter falls on to a black hole  from start to end. 
Despite that only a handful events are known so far and mostly were found
in the declining phase with sparsely sampled data,
a rich variety of the demography and physical processes involved have been revealed.
Recent discoveries of a few jetted events by Swift and MAXI indicated 
that relativistic jets can be launched in at least some of the TDEs \cite{bur11}. 
A distinctive X-ray lightcurve with large drops and recurrences 
was also seen in one case, which was interpreted as
 perhaps the most accessible signature of quiescent binary MBHs, via
 tidally disrupting a star \cite{liu14}.
Theoretical modeling also predicts that the effects of General Relativity may be tested, 
and the black hole spins be measured, from the well-sampled X-ray lightcurves of 
TDEs \cite{Kesden12}.
Moreover, TDEs provide an effective way to discover intermediate-mass black holes 
(IMBHs) with masses of $10^{2-5}$\,M$_{\odot}$ lurking at the centre of dwarf galaxies,
which are otherwise hard to find in observations.
 
EP is an ideal instrument to search systematically  for and study TDEs, and
is expected to detect them at the peaks of their X-ray flares out to 
at least a few hundred Mpc.
The estimated detection rate for EP may range from several tens to hundreds per year.
The majority can be caught at the rising phase of the flares, making it possible to
observe  in multiple wavelengths the complete process of TDEs from the onset of disruption. 
EP will also be able to detect more relativistic TDEs with jets  out to redshifts $z >1$, 
and more events with the possible signatures of binary MBHs. 
EP is expected to revolutionise the field of TDE research by 
detecting and characterising TDEs in large numbers with well-sampled data 
and catching them at an early flaring phase. 
This will greatly advance our understanding of
the demography,  formation and evolution of MBHs, as well as the physics of accretion
and jet formation. 

Moreover, EP will  discover new stellar- and intermediate-mass black holes lurking 
in our Milky Way and in nearby galaxies, 
in the galactic plane or even at the centres of globular clusters,
by detecting their X-ray outbursts due to some kind of instability of gas accretion. 

\subsection{Detect and precisely locate electromagnetic sources of gravitational-wave transients}

Direct detection of the long-sought gravitational wave (GW) signal is  
perhaps one of the most important events in modern physics and astrophysics, 
and will close the last missing link of the predictions of General Relativity. 
With the advent of the next generation of GW detectors (such as Advanced LIGO and Virgo), 
the detection of GW sources is highly anticipated towards the end of this decade.
Those detectors are expected to detect mergers of binary neutron stars (NS) or 
a neutron star with a black hole, that are strong transient GW sources, 
out to several-hundred Mpc.
The predicted combined detection rate, albeit with a considerable uncertainty, 
would be the order of $\sim$0.2--200\,yr$^{-1}$ from 2019 \cite{aasi13}.
However, GW sources are hard to locate, 
such that only ~10-30\% will have position accuracies better than 20 square degrees. 
The detection of  accompanying  electromagnetic  transients (EM counterparts)
is essential for  locating precisely and identifying  associated astrophysical sources, 
and for measuring redshifts (thus the source distance and  EM energy budget).
These are important data for understanding the nature of the GW events. 
Therefore the EM counterparts are of great interest to both the GW and 
astrophysical communities. 
A potentially even more far-reaching application is in cosmology;
joint detections of GW and EM events at high redshifts can be used 
to probe the geometry and expansion of the Universe,
since GW events from binary mergers are the standard sirens from which 
the luminosity distances could be inferred independently \cite{schutz86},
while the EM measurements lead to the redshifts.

NS-NS mergers are thought to be the origin of  short GRBs, and simultaneous detection of an SGRB-like X-ray transient with a GW transient would provide the definitive confirmation of this model.
In addition, recent models also predict that, in the case where a massive magnetar 
formed as an (intermediate) merger product, transient X-ray emission 
can be produced nearly isotropically on relatively short timescales up to hours \cite{zhang13};
the isotropy of the emission implies that the chance of being detected  is significantly
higher than that of highly beamed jet emission from short GRBs. 

With the combined high sensitivity, large FoV, manoeuvrability and rapid alert downlink,  
EP is expected to be a major contributor to this exciting new field.
By synergy with the next generation of GW detectors,
EP is expected to either detect the photonic sources simultaneously with GW events, or 
 rapidly observe the sky region covering the large error boxes provided by the GW detectors.

\subsection{Systematic surveys of X-ray transients}

EP will carry out systematic all-sky surveys to discover high-energy transients 
of various types over a wide range of timescales and at high cadence in the soft X-rays. 
It will also perform immediate follow-up observations of newly discovered transients 
with its narrow-field X-ray telescope onboard, 
and will issue fast alerts to trigger follow-up observations 
by  multi-wavelength facilities world-wide. 
Of particular interest, EP aims at detecting more high-redshift GRBs
and at higher redshifts than the currently known sample (7 GRBs with $z$=6.2--9.4).
These valued events, produced by stars in the early Universe, 
carry unique  information on early star-formation and metallicity evolution,
and hopefully on the first generation of stars and the re-ionization in the dark early Universe,
which are otherwise almost inaccessible from ground-based facilities.
Shock breakout emission from supernovae is 
the prompt X-ray emission produced as the outward-propagating shocks generated by
core-collapse breaks out of the stellar surface.
X-ray observations can yield important clues to the properties of the progenitor stars.
There are only a few candidate events known so far \cite{sod08} 
due to their elusiveness given the short burst durations ($\sim10^3$\,s) and moderate brightness.
EP is also expected to detect  more of such events. 
Other transient sources to detect and characterise  in large numbers  include
X-ray flashes, low-luminosity GRBs, X-ray rich GRBs,  GRB precursors, magnetars, 
stellar corona flares, classical novae, supergiant fast X-ray transients, 
and outbursts of active galactic nuclei and blazars, etc.

\section{Instruments}
EP will carry two scientific instruments---a survey instrument  Wide-field X-ray Telescope (WXT) 
with a large instantaneous Field-of-View (FoV, $60^\circ \times60^\circ$)
and a narrow-field ($1^\circ \times1^\circ$) Follow-up X-ray Telescope (FXT),
as well as a fast alert downlink system. 
To achieve both wide FoV and X-ray focusing, the novel Micro-Pore Optics (MPO) 
in the lobster-eye configuration \cite{angel79}\cite{dick}
is adopted for WXT. 
The MPO lobster-eye optics is made of a thin spherical micro-channel plate with millions of 
square micro-pores, the axes of which all point radially to a common centre of curvature. 
Grazing incidence reflection from the sides of the pores can focus incoming X-rays  
onto a focal sphere with a radius of half the curvature of the optics.
The point spread function, which is cruciform, has a central peaking spot 
produced by two reflections from adjacent walls of the pores and 
cross-arms produced by single reflections. 
Such an optics can yield a large focusing gain of $\sim$2000, 
and a moderate resolution of FWHM the order of arc-minutes  delivered by 
the currently available MPO pieces.
In practice, the MPO Lobster-eye optics can achieve a FoV 
of thousands of square degrees with very light weight,  
which is unique for wide-field X-ray imaging.

\begin{figure}
     \includegraphics[width=0.59\textwidth]{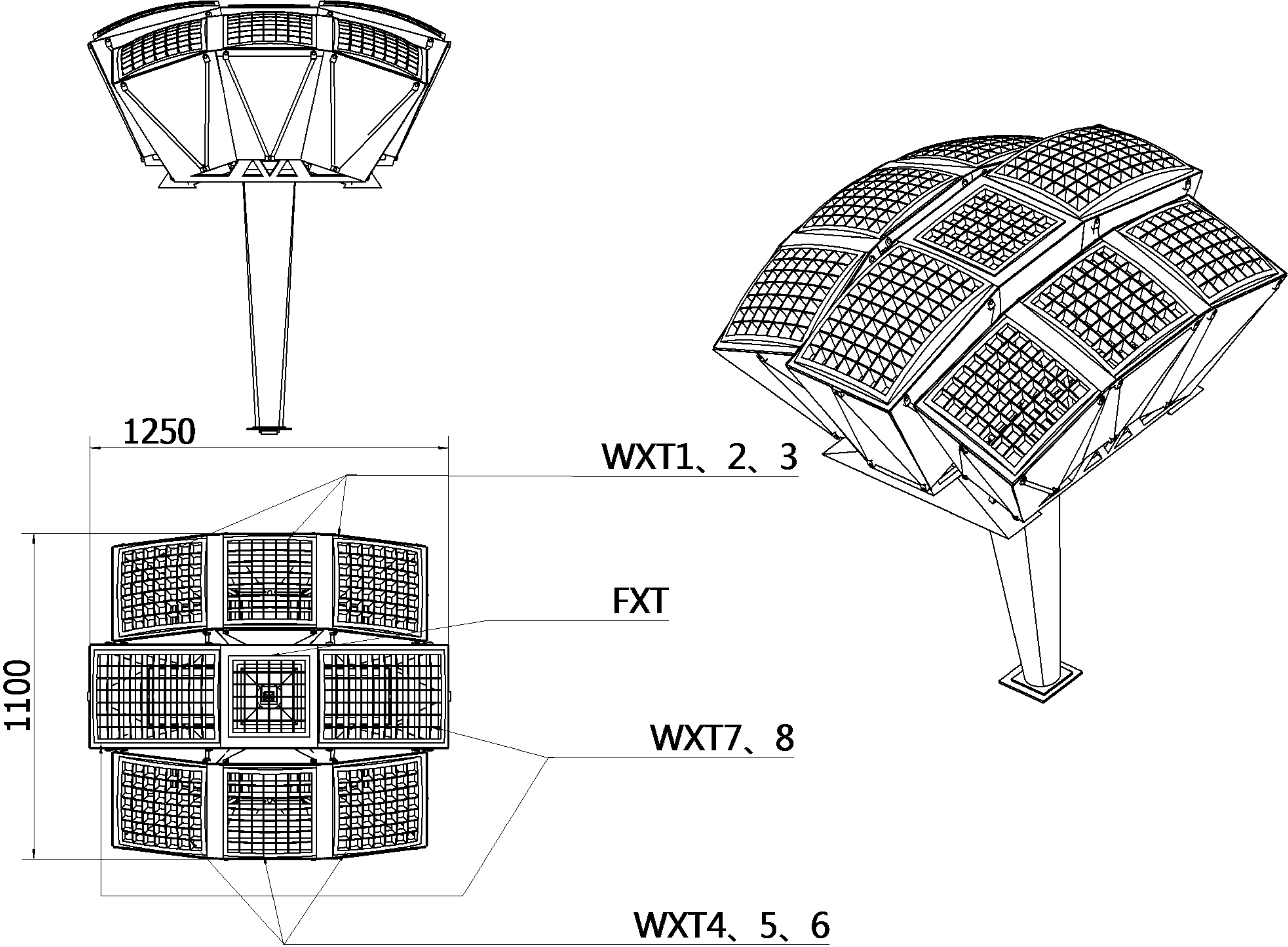}
	\includegraphics[width=0.39\textwidth]{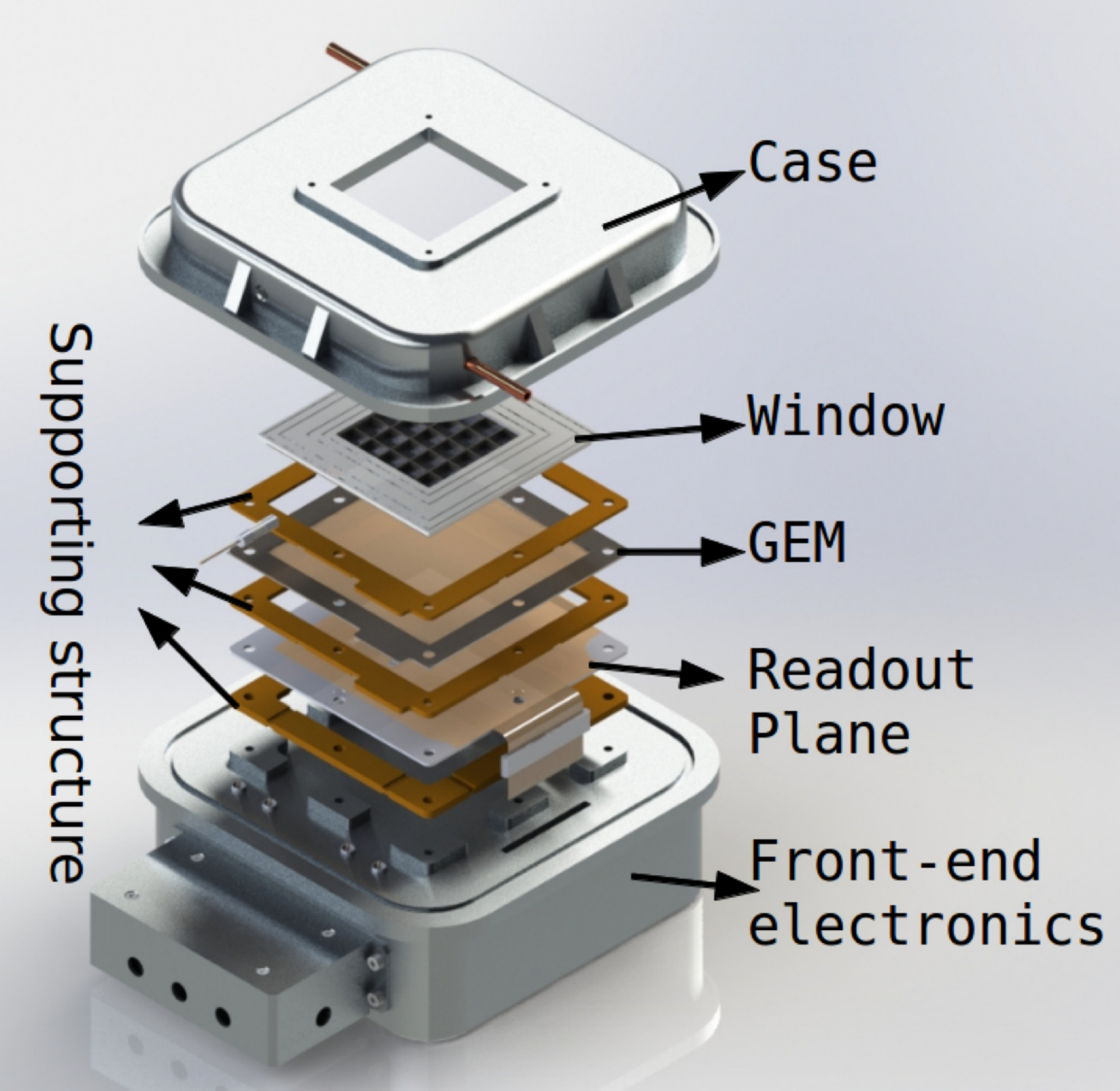}	
     \caption{(left) Layout of WXT modules and FXT. (right) Sketch of a GEM-based gas detector as the focal plane detector for      a WXT module.}
     \label{fig:layout}
     \end{figure}

WXT consists of eight modules with a 375\,mm focal length (Figure\,\ref{fig:layout}, left panel).  
Modules 1--6 are identical; each has an aperture size of 280\,mm$\times$280\,mm 
and is composed of $7\times7$ mosaicking MPO pieces, 
subtending $20^\circ \times20^\circ$ square degrees.
Modules 7 and 8 have an aperture size of 400\,mm$\times$280\,mm 
each and subtend $30^\circ \times20^\circ$ square degrees per module.
The total FoV of WXT is $60^\circ \times60^\circ$ ($\sim$1.1\,steradian).
To match the large-format focal plane,
gas detectors based on GEM (Gas Electron Multiplier) are under development, 
and Xenon gas will be used.
The detectors have a size of 14 $\times$ 14 cm$^2$ for each of modules 1--6,
and 14 $\times$ 20 cm$^2$ for modules 7--8.
Figure \ref{fig:layout} (right panel) shows a prototype of 
the gas detectors of one WXT module for preliminary tests.
The nominal bandpass of WXT is 0.5--4.0\,keV.
 

\begin{table}[]
\begin{center}
{\small \footnotesize 
\caption{\label{tab:specification} Specifications of WXT and FXT}
\begin{tabular}{lcc}\hline\hline
\noalign{\smallskip}
Parameters & WXT & FXT  \\ \hline    
\noalign{\smallskip}
Number of modules & 8 & 1\\
Field-of-view & $60^\circ \times60^\circ$ & $1^\circ \times1^\circ$ \\   
Focal length (mm) & 375 & 1,400 \\
Angular resolution FWHM (arcmin) & $<$5 & $<$5 \\
Bandpass (keV) & 0.5--4.0 & 0.5--4.0 \\
Energy resolution @1\,keV & 40\% & 100\,eV \\
Effective area (central focus) (cm$^2$)&  3 @0.7\,keV & 60  @1\,keV \\ 
Sensitivity (erg\,s$^{-1}$\,cm$^{-2}$ @1,000\,s) & $\sim1\times10^{-11}$& $\sim3\times10^{-12}$ \\
\noalign{\smallskip}\hline\\
\end{tabular}\\
}
\end{center}
\end{table}

Mounted at the centre of the payload, FXT adopts the same MPO focusing technology 
considering the simplicity of the technology and the relatively low budget of the mission.
It has a much longer focal length of 1.4m, 
leading to a much larger effective area than that of WXT
($\sim60$\,cm$^2$ @1\,keV). 
FXT has an aperture size of about 240\,mm$\times$240\,mm, 
which is mosaicked by $6\times6$ MPO pieces. 
A CCD will be used as the focal plane detector of FXT to gain better spectral resolution.

\begin{figure}
\centering
\includegraphics[width=0.47\textwidth]{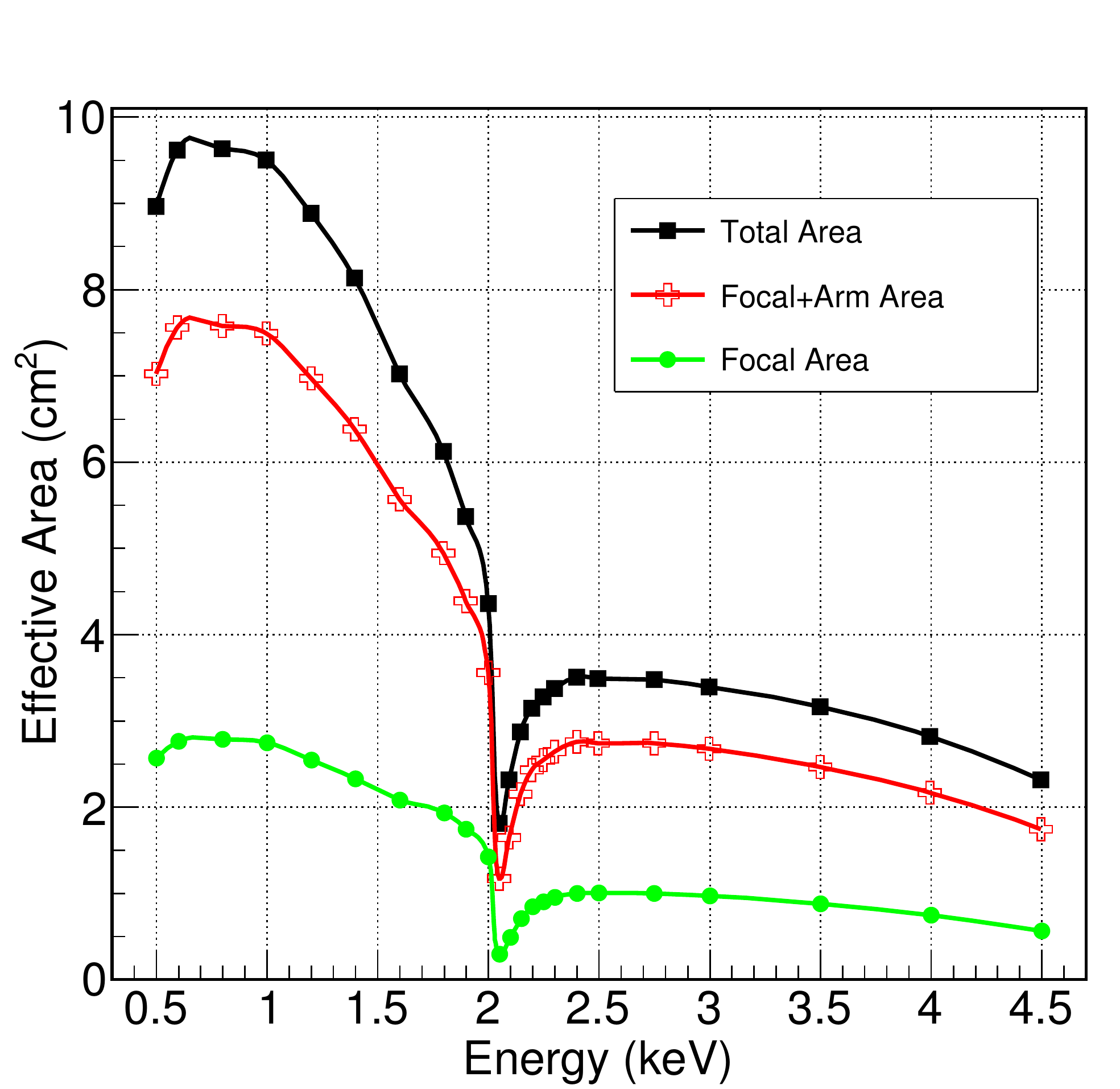}  
\includegraphics[width=0.49\textwidth]{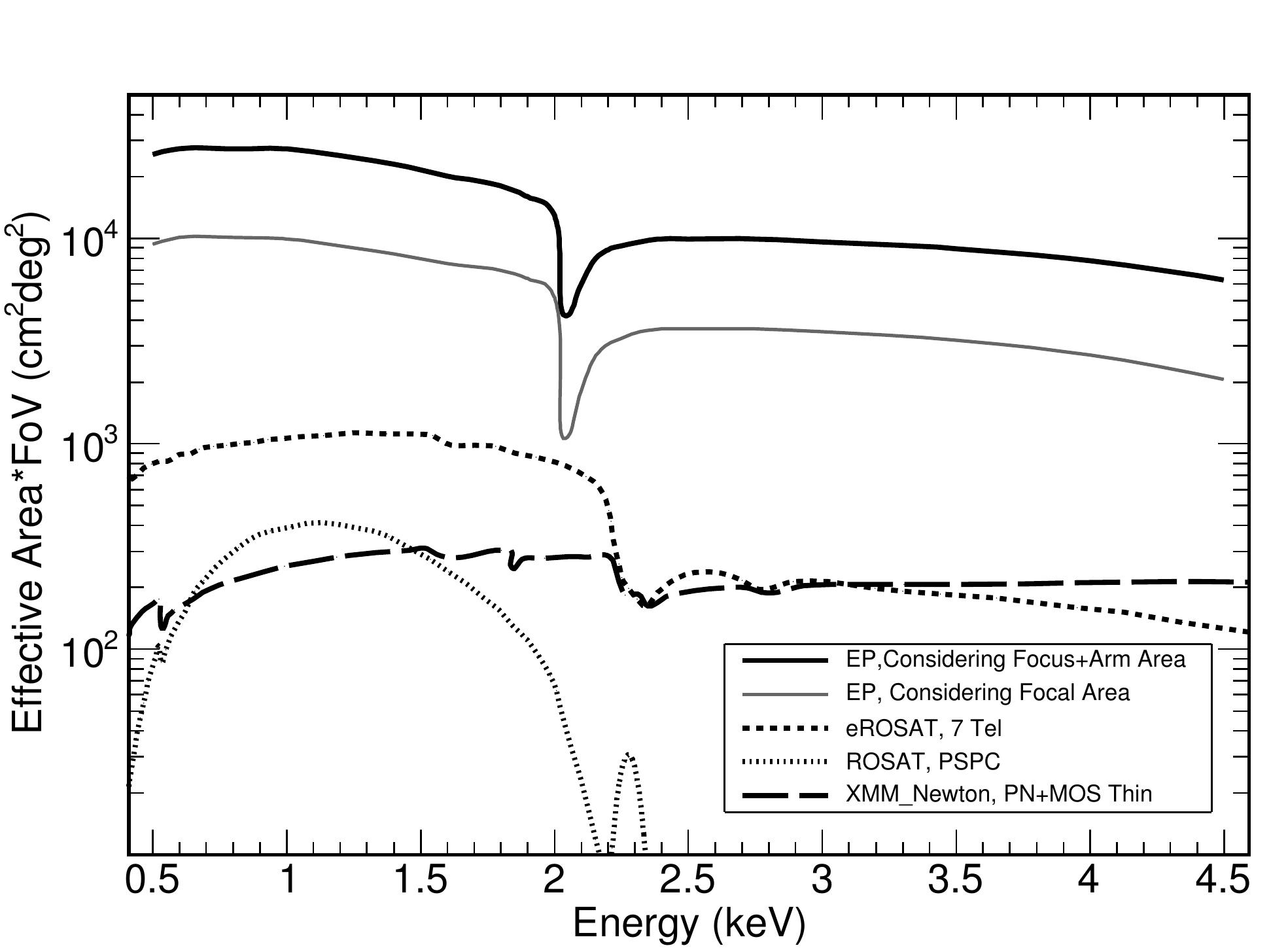}
\caption{(left) Effective area curves for EP/WXT with GEM detectors, 
for the central focal spot (green),  central plus the cruciform arms (red), and 
total (black; plus unfocused X-rays as diffuse background). 
The MPO arrays are coated with Iridium, and have surface roughness of 
$\sim$0.55\,nm and the tilts of pores following a Gaussian distribution
with $\sigma$=0.85\,arcmin.
The detector is filled with  Xenon gas, 
and has a thin window of a 40\,nm-thick Si$_{3}$N$_{4}$ foil coated with 30\,nm-thick Aluminum.     
(right) Grasp of WXT, in comparison with other focusing X-ray instruments.
Figures adopted from Zhao et al.\ (2014) \cite{zhao14}.}
\label{fig:effarea}
\end{figure}

The specifications of WXT and FXT are summarised in Table\,\ref{tab:specification}.
Figure\,\ref{fig:effarea} (left panel) shows the effective area curves for WXT.
These data are obtained via realistic ray-tracing simulations taking into account the imperfectness
of the MPO arrays \cite{zhao14}, which are in good agreement with the simulation results
from the Q software developed at the University of Leicester \cite{dick}. The Grasp parameter
(effective area times FoV)
is also shown in Figure\,\ref{fig:effarea} (right panel). 

\section{Mission Profile}

\begin{figure}
\begin{center}
     \includegraphics[width=0.55\textwidth]{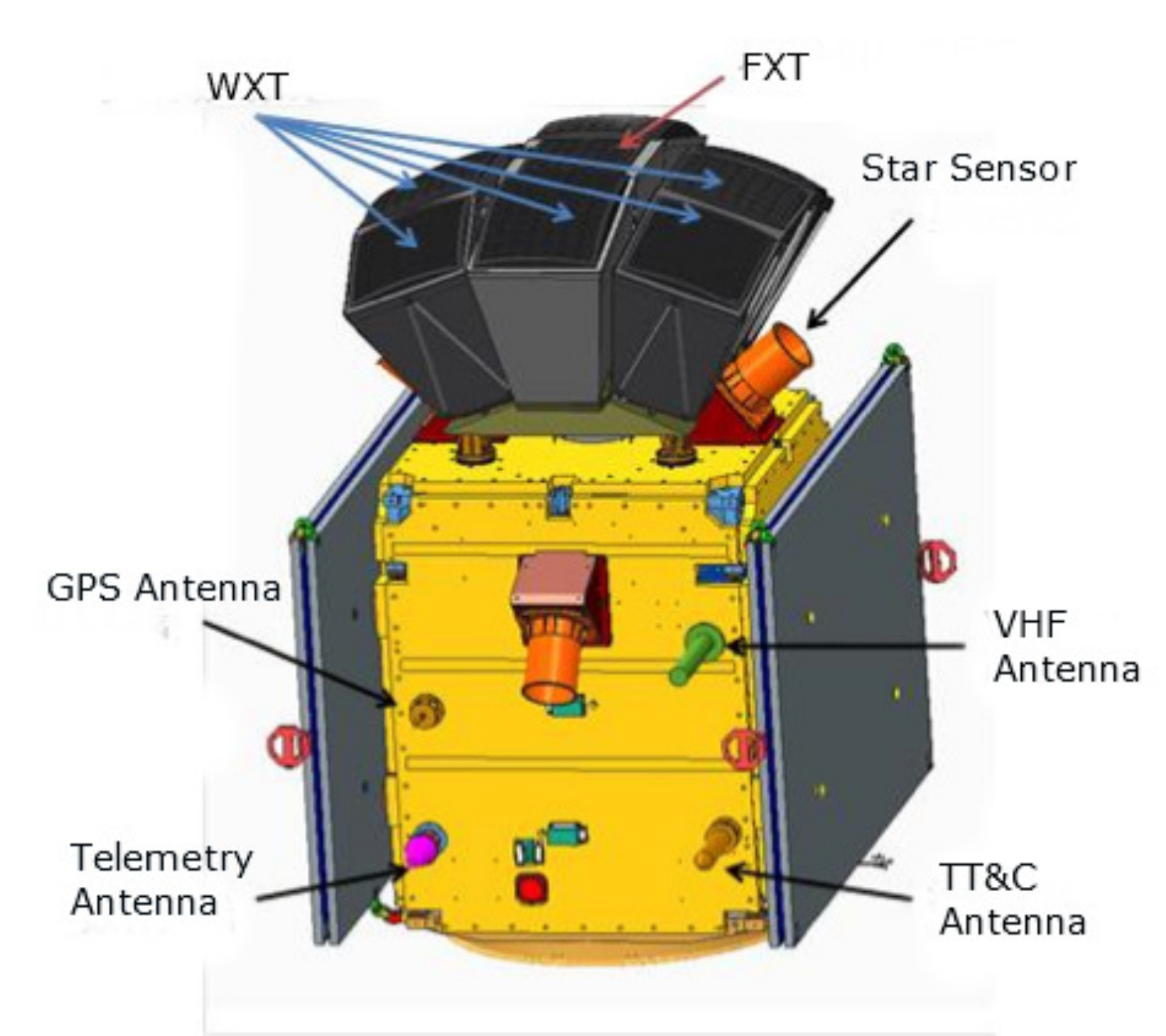}
     \includegraphics[width=0.44\textwidth]{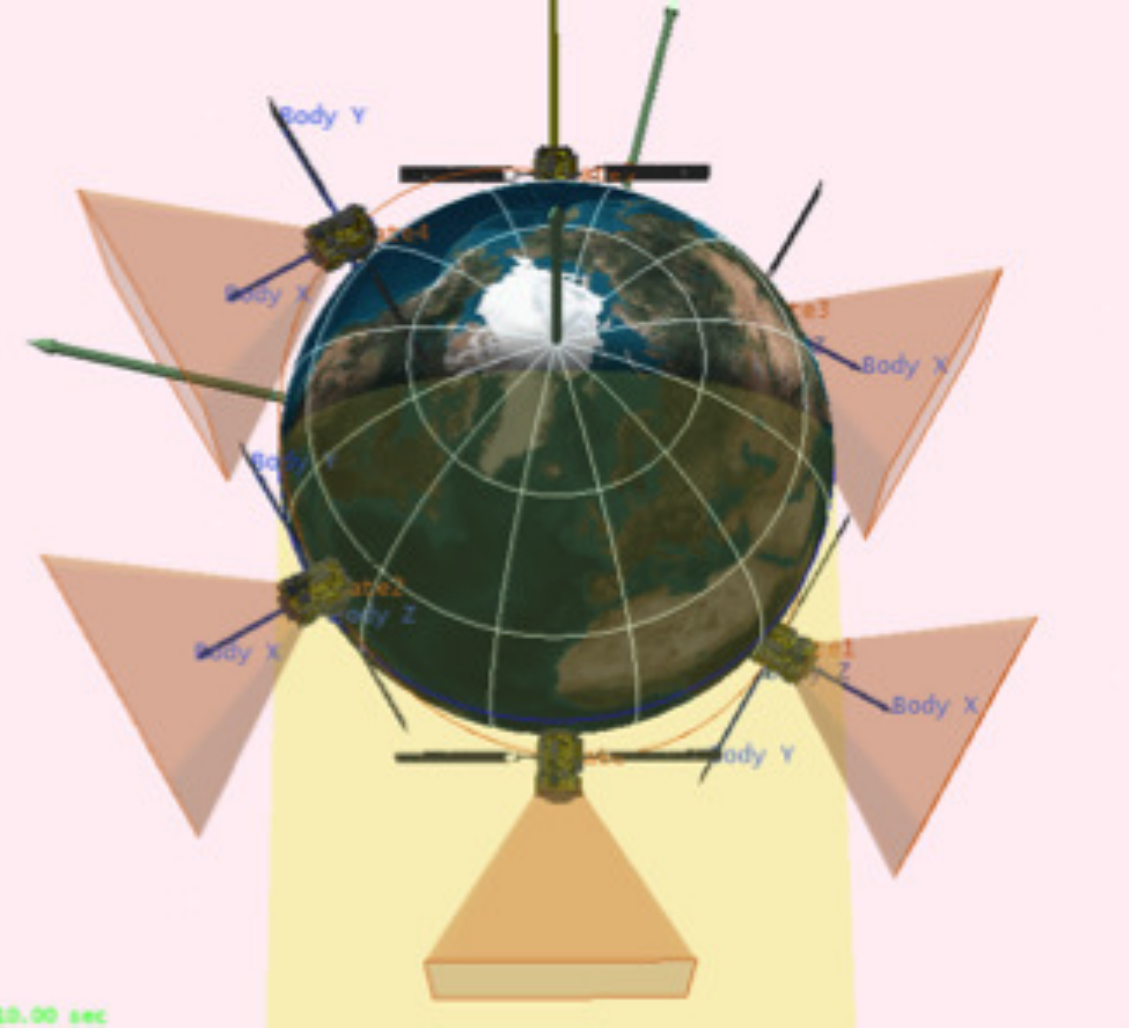}
          \caption{(left) Layout of the Einstein Probe satellite. (right) Illustration of the field-of-view and pointed observations in one orbit.}
     \label{fig:satellite}
     \end{center}
     \end{figure}

The combination of an array of wide-field lobster-eye modules to act as a soft X-ray transient monitor and a narrow-field lobster eye telescope for follow-up observations gives unprecedented grasp and sensitivity which will revolutionise X-ray transient astronomy.
The payload has low weight (150\,kg) and low power consumption (200W), which can be easily accommodated by one of the micro or small  satellite platforms readily available in China. 
One possible configuration is shown in Figure\,\ref{fig:satellite},
which has the total weight of the satellite about 450\,kg including the payload.
The satellite will be in a circular orbit at an altitude of $\sim$600 km and a period of 
$\sim$97 minutes, and an inclination angle of 30$^\circ$ or less. 

The survey strategy of EP will be composed of 
a series of pointings to mosaic the night sky in the directions avoiding the Sun.
During each 97-minute orbit of the satellite, five fields will be observed on the night-side of the sky, each with an 11-minute exposure. 
Over three orbits almost the entire night sky would be sampled, with cadences ranging from 5 to 25 times per day. 
Alternative schemes of e.g.\ $\sim20$ minutes exposure for each pointing are also under consideration.
The pointing directions are shifted by about 1 degree per day to compensate the daily movement of the Sun on the sky. In this way, the entire sky will be covered within half-a-year's operation.

Once a transient source is detected with WXT
and is classified and triggered by the processing and alerting system onboard, 
the satellite will slew to a new position to enable  pointed follow-up observations 
with FXT by targeting the new source within its FoV. 
Meanwhile, WXT continues to monitor  the new sky region centering the position of the transient.
It is essential to transmit alerts quickly to the ground, preferably within one minute or so.
There are two possibilities currently under consideration:
one is to make use of the French VHF ground station system 
by collaboration with CNES, 
and an alternative is to make use of the Chinese relay satellite network.

\section{Concluding remarks}

The X-ray Universe is rich in a large variety of energetic transients and variability, which
are highly relevant to some of the key questions in astrophysics.
The answers to these questions would have profound significance in 
 understanding our transient  Universe and its underlying  physical laws, 
and would  even revolutionise the relevant research. 
The Einstein Probe mission will attempt to address these questions by capturing faint flashes of X-ray radiation produced by energetic events within a cosmic horizon far beyond the reach of any current and previous missions. 
Its scientific impact will span a wide range of topics in astrophysics  and  fundamental physics, 
from  stars, compact objects in our Galaxy and in nearby galaxies, black holes, supernovae, GRBs, 
galaxies to cosmology.

\acknowledgments{
WY is grateful to many colleagues for valuable suggestions and discussion, and
special thanks are given to M. Matsuoka, S. Komossa, N. Gehrels, M. Feroci, L. Piro, B. Zhang, 
B. Cordier, N. Kawai.
The science consortium of Einstein Probe, composed of a large number of colleagues, 
is greatly acknowledged. 
This work is supported by the Strategic Priority Research Programme in Space Science of the CAS, Grant No. XDA04061100.}

\end{document}